\documentclass[aps,prd,twocolumn,nofootinbib,showpacs]{revtex4-1}

\usepackage{booktabs,makecell,multirow}
\usepackage{ulem}

\usepackage{graphicx,epsfig}
\usepackage{subfigure}

\usepackage[colorlinks=true, linkcolor=blue, citecolor=blue, urlcolor=blue]{hyperref}

\begin{document}

\title{A new analysis of the pQCD contributions to the electroweak parameter $\rho$ using the single-scale approach of principle of maximum conformality}

\author{Qing Yu}
\email[email:]{yuq@cqu.edu.cn}

\author{Hua Zhou}
\email[email:]{zhouhua@cqu.edu.cn}

\author{Jiang Yan}
\email[email:]{yjiang@cqu.edu.cn}

\author{Xu-Dong Huang}
\email[email:]{hxud@cqu.edu.cn}

\author{Xing-Gang Wu}
\email[email:]{wuxg@cqu.edu.cn}

\address{Department of Physics, Chongqing University, Chongqing 401331, P.R. China}
\address{Chongqing Key Laboratory for Strongly Coupled Physics, Chongqing 401331, P.R. China}

\date{\today}

\begin{abstract}

It has been observed that conventional renormalization scheme and scale ambiguities for the pQCD predictions can be eliminated by using the principle of maximum conformality (PMC). However, being the intrinsic nature of any perturbative theory, there are still two types of residual scale dependences due to uncalculated higher-order terms. In the paper, as a step forward of our previous work [Phys.Rev.D {\bf 89},116001(2014)], we reanalyze the electroweak $\rho$ parameter by using the PMC single-scale approach. Using the PMC conformal series and the Pad$\acute{e}$ approximation approach, we observe that the residual scale dependence can be greatly suppressed and then a more precise pQCD prediction up to ${\rm N^4LO}$-level can be achieved, e.g. $\Delta\rho|_{\rm PMC}\simeq(8.204\pm0.012)\times10^{-3}$, where the errors are squared averages of those from unknown higher-order terms and $\Delta\alpha_s(M_Z)=\pm 0.0010$. We then predict the magnitudes of the uncertainties of the $W$-boson mass and the effective leptonic weak-mixing angle: $\delta M_{W}|_{\rm N^4LO} =-0.26$ MeV and $\delta \sin^2{\theta}_{\rm eff}|_{\rm N^4LO}=0.14\times10^{-5}$, which are well below the precision anticipated for the future electron-position colliders such as FCC, CEPC and ILC. Thus by measuring those parameters, it is possible to test SM with high precision.

\end{abstract}

\maketitle

As is well-known, the electroweak $\rho$ parameter~\cite{Veltman:1977kh} is one of the most important parameters for standard model (SM). It is defined as the ratio between the strengths for the charged and neutral currents. At the Born level, $\rho=1$, and its QCD correction $\Delta\rho$ can be derived from the transverse parts of $W$ and $Z$ boson self-energies at the zero momentum transfer~\cite{Djouadi:1987di}:
\begin{equation}
\Delta\rho={\prod_Z(0)\over M^2_Z}-{\prod_W(0)\over M^2_W}.
\end{equation}

At the present, the correction $\Delta\rho$ has been calculated up to four-loop level~\cite{Veltman:1977kh, Djouadi:1987di, Djouadi:1987gn, Kniehl:1988ie, Avdeev:1994db, Chetyrkin:1995ix, Chetyrkin:1995js, Chetyrkin:2006bj, Schroder:2005db, Boughezal:2006xk, Faisst:2006sr}. The fixed-order pQCD predictions are usually assumed to suffer from an uncertainty in fixing the renormalization scale ($\mu_r$). This ambiguity in making fixed-order predictions occurs because one usually assumes an arbitrary renormalization scale, (representing a typical momentum flow of the process which is assumed to be the effective virtuality of the strong interaction in that process), together with an arbitrary range to ascertain its uncertainty. For the parameter $\Delta\rho$, its renormalization scale $\mu_r$ is usually chosen as $M_t$ and then varying it within the range of $[{M_t}/2, 2 M_t]$ to ascertain its uncertainty. However, this {\it ad hoc} assignment of the renormalization scale causes the coefficients of the QCD running coupling at each order to be strongly dependent on the choice of $\mu_r$ as well as the renormalization scheme. Moreover, we do not know how wide a range the renormalization scale and scheme parameters should vary in order to achieve reasonable predictions of their errors. Thus the renormalization scale uncertainty becomes one of the most important systematic errors for usual pQCD predictions.

Such error is however unnecessary and could be removed. In year 2014, we have successfully applied the multi-scale method of principle of maximum conformality (PMC)~\cite{Brodsky:2011ta, Brodsky:2011ig, Mojaza:2012mf, Brodsky:2013vpa} to eliminate the conventional renormalization scheme and scale ambiguity in the QCD correction $\Delta\rho$~\cite{Wang:2014wua}. The purpose of PMC is not to find an optimal renormalization scale but to determine the effective coupling constant (whose argument is called as the PMC scale) with the help of renormalization group equation (RGE)~\cite{Wu:2013ei, Wu:2014iba, Wu:2019mky}, and its prediction satisfies the requirement of the renormalization group invariance~\cite{Brodsky:2012ms}. In Ref.\cite{Wang:2014wua}, we have found that by applying the PMC, the conventional renormalization scheme and scale ambiguities can be eliminated, and because of eliminating divergent renormalon terms~\cite{Beneke:1994qe, Neubert:1994vb, Beneke:1998ui}, the convergence of the pQCD series can also be greatly improved, leading to a more reliable and accurate pQCD prediction. More explicitly, it has been found that under conventional scale-setting approach, by defining a ratio $\Delta R= (\Delta\rho/3X_t -1)$ with $X_{t}=(G_{F}M^{2}_{t})/(8\sqrt{2}\pi^{2})$, the renormalization scale error of $\Delta R$ is $\sim \pm9 \%$ at the two-loop level, $\sim\pm4\%$ at the three-loop level, and $\sim \pm 2.5\%$ at the four-loop level for ${\mu_r} \in[{M_t}/2, 2 M_t]$, respectively; These facts explains why the conventional scale uncertainty constitutes an important error for estimating the $\rho$ parameter. On the other hand, by applying the PMC, the four-loop estimation $\Delta\rho|_{\rm N^3LO}$ is fixed to be $\sim 8.2\times10^{-3}$ for any choice of $\mu_r$, in which $\mu_r$ only needs to be in perturbative region ($\gg\Lambda_{\rm QCD}$) to ensure the pQCD calculation.

For the PMC multi-scale method~\cite{Brodsky:2011ta, Brodsky:2011ig, Mojaza:2012mf, Brodsky:2013vpa}, we need to absorb different types of $\{\beta_i\}$-terms into the corresponding $\alpha_s$ via an order-by-order manner. Different types of $\{\beta_i\}$-terms as determined from the RGE lead to different running behaviors of $\alpha_s$ at different orders, and hence, determine distinct PMC scales at each order. Becomes the PMC scales are of perturbative nature, the precision of the PMC scale for higher-order terms decreases with the increment of perturbative orders due to the less $\{\beta_i\}$-terms are known in those higher-order terms. The PMC has thus two kinds of residual scale dependence due to the unknown perturbative terms~\cite{Zheng:2013uja}, i.e. the last terms of the PMC scales are unknown (\textit{first kind of residual scale dependence}) and the last terms of the pQCD approximant are unfixed since its PMC scale cannot be determined (\textit{second kind of residual scale dependence}). Generally, those two residual scale dependence suffer from both the $\alpha_s$-power suppression and the exponential suppression, but could be large due to poor pQCD convergence of the perturbative series of either the PMC scale or the pQCD approximant. A detailed discussion on residual scale dependence can be found in the review~\cite{Wu:2019mky}.

In year 2017, we have suggested a single-scale method for PMC which effectively suppresses those residual scale dependence, e.g. the first kind of residual scale dependence is greatly suppressed by using an overall PMC scale with the same precision for all orders and the second one is exactly removed~\cite{Shen:2017pdu}. The PMC single-scale method is equivalent to multi-scale method in the sense of perturbative theory, and it replaces the individual PMC scales at each order by an overall scale, which effectively replaces those individual PMC scales derived under the PMC multi-scale method in the sense of a mean value theorem. The PMC single scale is independent to any choice of renormalization scale and it can be regarded as the overall effective (physical) momentum flow of the process. It has been demonstrated that the PMC single-scale prediction is scheme independent up to any fixed order~\cite{Wu:2018cmb}, satisfying the standard renormalization group invariance. In the following, we shall adopt the PMC single-scale method to analyze the shift $\Delta\rho$.

As done in Ref~\cite{Wang:2014wua}, after accomplishing the transformation of the pQCD series over the top-quark running-mass $m_t$ given in Refs.\cite{Chetyrkin:2006bj, Schroder:2005db, Boughezal:2006xk, Faisst:2006sr} into the series over the pole-mass $M_t$~\footnote{Because only those $\{\beta_i\}$-terms that are pertained to RGE of $\alpha_s$ should be applied for determining the $\alpha_s$-running behavior, and to avoid the confusion of using the $\{\beta_{i}\}$-terms pertaining to the renormalization of $m_t$, we do this transformation.}, the $\Delta\rho$ parameter up to four-loop QCD corrections (${\rm N^3LO}$) can be expressed as
\begin{eqnarray}
\Delta\rho&=&3X_{t}\big[1-11.4396a_s(\mu_r)+\big(-404.981-125.836l_t \nonumber\\
&&+n_f(28.5794+7.62643l_t)\big)a^2_s(\mu_r)+\big(-20372.1 \nonumber\\
&&-9537.9l_t-1384.2l_t^2+n_f(2843.09+1223.87l_t \nonumber\\
&&+167.782l_t^2)+n^2_f(-73.5582-38.1059l_t \nonumber\\
&&-5.08429l_t^2)\big)a^3_s(\mu_r)\big],
\label{deltarho}
\end{eqnarray}
where $a_{s}(\mu_{r})=\alpha_{s}(\mu_{r})/4\pi$, $X_{t}=(G_{F}M^{2}_{t})/(8\sqrt{2}\pi^{2})$, $G_{F}$ is the Fermi constant, $n_f$ is the active light-flavor numbers, and $l_t=\ln({\mu^2_r}/{M^2_t})$. By using the general degeneracy relations in QCD~\cite{Bi:2015wea} among different orders, $\Delta\rho$ can be further rewritten as
\begin{eqnarray}
\Delta\rho&=&3X_{t}\left[1+\sum^{3}_{i=1}r_i(\mu_{r})a^i_{s}(\mu_{r})\right] \label{convser} \\
&=&3X_{t}\big[1+r_{1,0}(\mu_{r})a_{s}(\mu_{r}) + \big( r_{2,0}(\mu_{r})  + \beta_{0}r_{2,1}(\mu_{r})\big) \nonumber\\
&& a_{s}^{2}(\mu_{r})+\big( r_{3,0}(\mu_{r}) +\beta_{1}r_{2,1}(\mu_{r})+2\beta_{0}r_{3,1}(\mu_{r})\nonumber\\
&&+ \beta_{0}^{2}r_{3,2}(\mu_{r})\big) a_{s}^{3}(\mu_{r})\big],
\label{rhorij}
\end{eqnarray}
where the fist two $\{\beta_i\}$-functions $\beta_{0} = 11-{2\over 3}n_{f}$, $\beta_{1} = 102-{38\over 3} n_{f}$, $r_{i}$ and $r_{i,j}$ can be derived from Eq.(\ref{deltarho}). $r_{i,0}$ are scale-invariant conformal coefficients, and $r_{i, {j\neq 0}}$ are nonconformal coefficients, which are generally functions of $\ln({\mu^2_r}/{M_t^2})$
\begin{eqnarray}
r_{i,j} = \sum_{k=0}^{j} C_j^k \ln^k(\mu_r^2/M_t^2) \hat{r}_{i-k,j-k},
\label{rij}
\end{eqnarray}
where $C_j^k={j!}/{k!(j-k)!}$ and $\hat{r}_{i,j}=r_{i,j}|_{\mu_r=M_t}$.

Following the standard procedures of the PMC single-scale method, we obtain the following conformal series
\begin{eqnarray}
\Delta\rho &=& 3X_{t}\left(1+\hat{r}_{1,0}a_{s}(Q_{*}) + \hat{r}_{2,0} a_{s}^{2}(Q_{*})+  \hat{r}_{3,0}a_{s}^{3}(Q_{*}) \right),
\label{rhori0PMC}
\end{eqnarray}
in which the PMC scale $Q_{*}$ is independent to any choice of $\mu_r$, and because the conformal coefficients $\hat{r}_{i,0}$ are also scale invariant, the resultant pQCD series is thus free of renormalization scale ambiguity. At present, $Q_{*}$ can be determined up to next-to-leading-log (NLL) accuracy with the known ${\rm N^{3}LO}$ pQCD series, e.g.
\begin{eqnarray}
\ln{Q^2_{*}\over M^2_t}= T_0+T_1 a_s(M_t) +{\cal O}(a_s^2),
\label{rhoPMCscale}
\end{eqnarray}
where
\begin{eqnarray}
T_0 &=& -{\hat{r}_{2,1}\over \hat{r}_{1,0}},  \\
T_1 &=& {2(\hat{r}_{2,0}\hat{r}_{2,1}-\hat{r}_{1,0}\hat{r}_{3,1})\over  \hat{r}_{1,0}^2}  +{(\hat{r}_{2,1}^2-\hat{r}_{1,0}\hat{r}_{3,2})\over \hat{r}_{1,0}^2}\beta_0.
\end{eqnarray}

To do the numerical calculation, we take the parameter values from the Particle Data Group~\cite{PDG2020}: the $W$-boson mass $M_{W}=80.379\pm0.012$ GeV, the $Z^0$-boson mass $M_{Z}=91.1876\pm0.0021$ GeV, and the top-quark pole mass $M_{t}=172.76\pm0.30$ GeV, which corresponds to a recent determination of $\overline{\rm MS}$ mass $\overline{m}_t(\overline{m}_t)= 162.629$ GeV~\cite{Huang:2020rtx}. The Fermi constant $G_{F}=1.16638\times10^{-5}~{\rm GeV}^{-2}$. The four-loop $\alpha_s$ running behavior is adopted, whose asymptotic scale is fixed by $\alpha_s(M_{Z})=0.1179\pm0.0010$~\cite{PDG2020}, which gives $\Lambda_{{\rm QCD},n_f=5}=207^{+12}_{-11} $~MeV.

\begin{figure}[htb]
\includegraphics[width=0.45\textwidth]{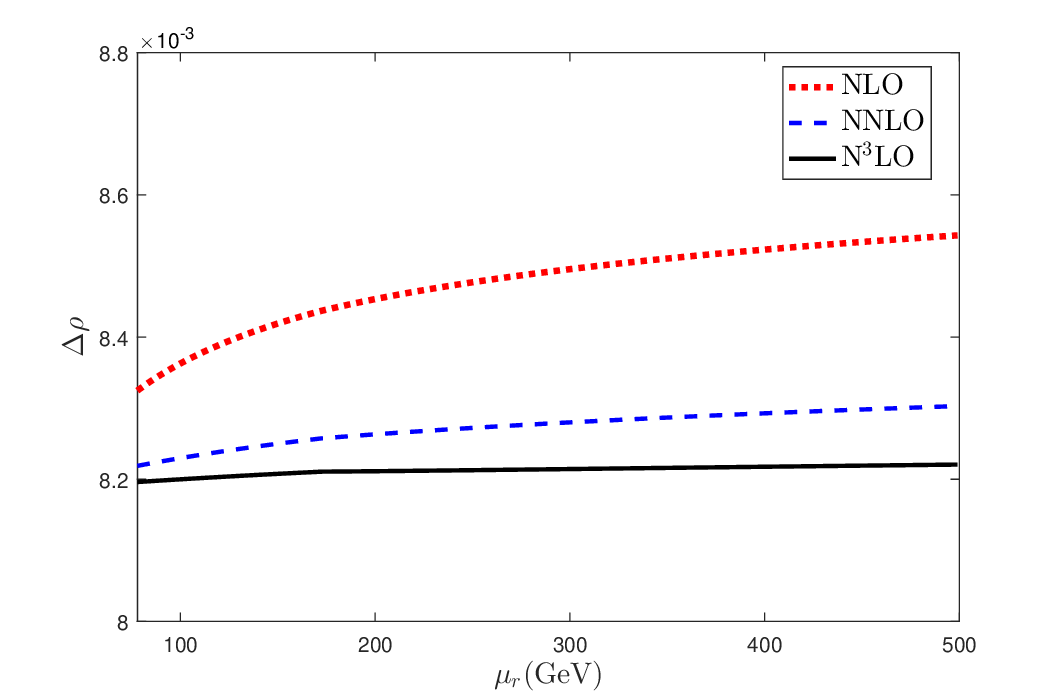}
\caption{The $\Delta\rho$ up to ${\rm N^3LO}$-level versus the renormalization scale $\mu_{r}$ under conventional scale-setting approach.}
\label{rhoconv}
\end{figure}


We present the QCD correction $\Delta\rho$ up to ${\rm N^3LO}$-level versus the renormalization scale $\mu_{r}$ under conventional and PMC single-scale approaches in Figs.(\ref{rhoconv}, \ref{rhopmc}), respectively. In agreement with conventional wisdom, Fig.~\ref{rhoconv} shows that the conventional $\mu_r$ dependence shall be greatly suppressed when more and more loop terms have been included. More explicitly, we have
\begin{eqnarray}
\Delta\rho|_{\rm Conv}&=& \left(9.353-0.916^{+0.072}_{-0.096}-0.180^{-0.021}_{+0.062} \right. \nonumber\\
& & \quad\quad\quad \left. -0.047^{-0.008}_{+0.021} \right) \times 10^{-3} \nonumber\\
&=& \left(8.211^{+0.043}_{-0.013}\right)\times10^{-3},
\label{convdeltarho}
\end{eqnarray}
where the central values are for $\mu_r=M_t$, and the errors are for $\mu_r\in[M_t/2, 2M_t]$, which are less than $1\%$. Such small net error is caused by cancellation of scale dependence among different orders, $\Delta\rho|_{\rm Conv}$ at each order still shows strong dependence on $\mu_r$. We define a factor $k_i$ to measure the scale
dependence of each loop terms under conventional scale-setting approach,
\begin{eqnarray}
k_i=\left| {\Delta\rho^{\rm N^{i}LO}|_{\mu_r=M_t/2}-\Delta\rho^{\rm N^iLO}|_{\mu_r=2M_t}\over\Delta\rho^{\rm N^iLO}|_{\mu_r=M_t}} \right|.
\end{eqnarray}
The magnitude of $k_i$ shows that there is high scale dependence for each loop terms, which are $\sim 18\%$, $\sim 46\%$ and $\sim 62\%$ for ${\rm NLO}$-terms, ${\rm N^2LO}$-terms and ${\rm N^3LO}$-terms, respectively. Though scale dependent, the conventional series shows a required pQCD convergence, e.g. the relative importance of the ${\rm LO}$-terms: ${\rm NLO}$-terms: ${\rm N^2LO}$-terms: ${\rm N^3LO}$-terms equal to $1: \left(-9.8^{+0.8}_{-1.0}\right)\%: \left(-1.9^{-0.2}_{+0.6}\right)\%: \left(-0.5^{-0.1}_{+0.2}\right)\%$, where the errors are for $\mu_{r}\in[M_{t}/2, 2M_{t}]$.

On the other hand, the PMC predictions are free of $\mu_r$; i.e. for any choice of $\mu_r$, we have
\begin{eqnarray}
\Delta\rho|_{\rm PMC} &\equiv& (9.353-1.266+0.087+0.030)\times10^{-3} \nonumber\\
                      &\equiv& 8.204\times10^{-3}.
\label{pmcnumberical}
\end{eqnarray}
It is noted that the relative importance of the ${\rm LO}$-terms: ${\rm NLO}$-terms: ${\rm N^2LO}$-terms: ${\rm N^3LO}$-terms are fixed to $1: -13.5\%: +0.9\%: +0.3\%$. It shows a better convergence than that of the conventional series, which is due to the elimination of divergent renormalon terms. The PMC scale $Q_{*}$ can be fixed up to NLL-accuracy, i.e
\begin{eqnarray}
Q_{*}=M_t \exp\left[-1.874 - 1.201 \alpha_s(M_t)  \right].
\end{eqnarray}

\begin{figure}[htb]
\includegraphics[width=0.45\textwidth]{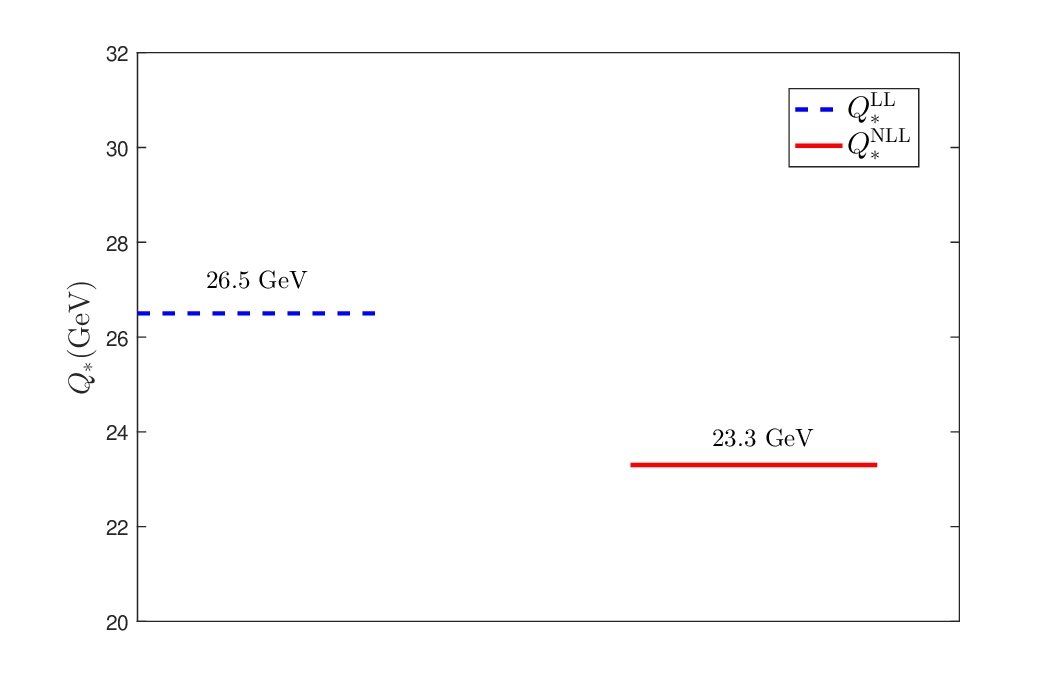}
\caption{The PMC scale $Q_*$ for $\Delta\rho$, where the superscripts ``LL" and ``NLL" are for LL and NLL accuracies, respectively.}
\label{Qstar}
\end{figure}

We present the PMC scale $Q_*$ up to LL-level and NLL-level in Fig.~\ref{Qstar}. $Q_*$ behaves convergently, e.g. the magnitude of the NLL-terms is only about $7\%$ of the LL-terms, and then the \textit{first kind of residual scale dependence} is greatly suppressed. Considering the perturbative nature of $Q_*$, as a conservative estimation of contribution of its unknown NNLL-term, we take the unknown NNLL-coefficient as the known NLL-coefficient, e.g. the NNLL-term is taken as $\pm 1.201\alpha_s^2(M_t)$, which leads to $Q^{\rm NNLL}_*=23.3\pm 0.3$ GeV. Then, we obtain
\begin{equation}
\Delta\rho|_{\rm PMC} = (8.204\pm0.003) \times 10^{-3}.
\end{equation}
Another more conservative way to estimate the \textit{first kind of residual scale dependence} is to set $Q^{\rm NNLL}_*=Q^{\rm NLL}_* \pm |Q^{\rm NLL}_*-Q^{\rm LL}_*|=23.3\pm 3.2$ GeV, which gives~\footnote{This shall finally leads to $\Delta'\rho|^{\rm High~order}_{\rm PMC} \simeq (^{+0.025}_{-0.029} )\times 10^{-3}$, which is comparable to that of conventional scale-setting. Such a much larger \textit{first kind of residual scale dependence} is reasonable, since slight change of PMC scale could lead to large effects due to the explicit breaking of conformal invariance~\cite{Brodsky:2012sz}. }
\begin{equation}
\Delta'\rho|_{\rm PMC} = (8.204^{+0.024}_{-0.029}) \times 10^{-3}.
\end{equation}

An important issue for pQCD theory is to know what are the contributions of unknown higher-order terms. In the PMC single-scale method, the unknown magnitude of the uncalculated even higher-order terms is called as the newly second kind of residual scale dependence~\cite{Huang:2021kzc}. In the literature, the Pad$\acute{e}$ approximation approach (PAA)~\cite{Basdevant:1972fe, Samuel:1992qg, Samuel:1995jc} is one of the approaches for estimating the unknown $(n+1)_{\rm th}$-order coefficient from a given $n_{\rm th}$-order perturbative series. It works effectively when we have known enough perturbative terms, and the renormalization scheme-and-scale independent conformal series is helpful for predicting the magnitudes of higher-order terms. Many successful applications by using PAA together with the PMC conformal series have been done in the literature~\cite{Du:2018dma, Yu:2020tri, Huang:2020rtx, Yu:2019mce, Yu:2018hgw, Huang:2021kzc}.

The idea of PAA is to predict the unknown $(n+1)_{\rm th}$-order coefficient by expanding a selected generating function which is fixed by matching to the known $n_{\rm th}$-order perturbative series. More explicitly, for a given pQCD series $\rho_n(Q)=\sum_{i=1}^{n} C_{i} a^{i}_s$, whose coefficients $C_{i}$ could be set as $\hat{r}_{i,0}$ for PMC series (\ref{convser}) or $r_i$ for conventional series (\ref{rhori0PMC}), the $[N/M]$-type generating function is defined as
\begin{equation}
\rho^{[N/M]}_n(Q) =  a_s \times \frac{b_0+b_1 a_s + \cdots + b_N a_s^N}{1 + c_1 a_s + \cdots + c_M a_s^M}, \label{eqrho1}
\end{equation}
where $M\geq 1$ and $N+M+1=n$. It can be expanded as a power series over $a_s$, i.e.,
\begin{equation}
\rho^{[N/M]}_n(Q) = \sum_{i=1}^{n} C_{i} a_s^{i} + C_{n+1}\; a_s^{n+1} + \cdots.
\label{eqrho2}
\end{equation}
Equating the above two equations (\ref{eqrho1}, \ref{eqrho2}) up to $n_{\rm th}$-order level, the coefficients $b_{i\in[0,N]}$ and $c_{i\in[1,M]}$ can be expressed by using the known coefficients $C_{i\in[1,n]}$. Thus the unknown $(n+1)_{\rm th}$-order coefficient $C_{n+1}$ can be expressed by $b_{i\in[0,N]}$ and $c_{i\in[1,M]}$, and hence by the known coefficients $\{C_{1},...,C_{n}\}$.

As for $\Delta\rho$, its coefficients up to ${\rm N^3LO}$ level are known, and the predicted ${\rm N^4LO}$-coefficient for conventional series (\ref{rhori0PMC}) is
\begin{eqnarray}
r^{\rm PAA}_{4}(\mu_r)|^{[1/1]} &=& \frac{{r^2_{3}(\mu_r)}}{{r_{2}(\mu_r)}},\nonumber\\
r^{\rm PAA}_{4}(\mu_r)|^{[0/2]} &=& \frac{2{r_{1}(\mu_r)}{r_{2}(\mu_r)}{r_{3}(\mu_r)}-{r^3_{2}(\mu_r)}}{{r^2_{1}(\mu_r)}}\nonumber\\
r^{\rm PAA}_{4}(\mu_r)|^{[1/1]} =&& (-243.926^{+169.762}_{-172.874})\times10^3\nonumber\\
r^{\rm PAA}_{4}(\mu_r)|^{[0/2]} =&& (-228.798^{+170.700}_{-180.500})\times10^3
\label{PAA11}
\end{eqnarray}
which is for the preferable diagonal [1/1]-type generating function for conventional series~\cite{Gardi:1996iq, Cvetic:1997qm}, and then our prediction of $\Delta\rho^{\rm N^4LO}$ is
\begin{eqnarray}
\Delta\rho^{\rm N^4LO}|^{[1/1]}_{\rm PAA+Conv}&=& \pm \left(0.012_{+0.003}^{-0.007}\right) \times 10^{-3},\nonumber\\
\Delta\rho^{\rm N^4LO}|^{[0/2]}_{\rm PAA+Conv}&=& \pm \left(0.011_{+0.003}^{-0.007}\right) \times 10^{-3},
\end{eqnarray}
whose central value is $\sim26\%$ of ${\rm N^3LO}$-terms for $\mu_r=M_t$, and the errors are for $\mu_r\in [M_t/2, 2M_t]$. The predicted ${\rm N^4LO}$-coefficient for PMC series (\ref{convser}) is
\begin{eqnarray}
\hat{r}^{\rm PAA}_{4,0}|^{[1/1]} &=& \frac{{\hat{r}^2_{3,0}}}{{\hat{r}_{2,0}}},\nonumber\\
\hat{r}^{\rm PAA}_{4,0}|^{[0/2]} &=& \frac{2{\hat{r}_{1,0}}{\hat{r}_{2,0}}{\hat{r}_{3,0}}-{\hat{r}^3_{2,0}}}{{\hat{r}^2_{1,0}}},\nonumber\\
r^{\rm PAA}_{4,0}(\mu_r)|^{[1/1]} &=& 55.701\times10^3\nonumber\\
r^{\rm PAA}_{4,0}(\mu_r)|^{[0/2]} &=& (-24.671)\times10^3
\label{PAA02}
\end{eqnarray}
which is derived by using the preferable $[0/2]$-type generating function for the PMC series~\cite{Du:2018dma}, and then our prediction of $\Delta\rho^{\rm N^4LO}$ is
\begin{eqnarray}
\Delta\rho^{\rm N^4LO}|_{\rm PAA+PMC} &=&\pm0.005\times10^{-3},
\label{pmcn4lo}
\end{eqnarray}
which is about $17\%$ of ${\rm N^3LO}$-term. To make an comparison, the calculated [1/1]-type prediction: $\Delta\rho^{\rm N^4LO}|^{[1/1]}_{\rm PAA+PMC} =\pm0.010\times10^{-3}$, which is about $34\%$ of ${\rm N^3LO}$-term.

The squared average of the above two errors could be treated as an estimator of how the uncalculated higher-order terms contribute to $\Delta\rho$,
\begin{eqnarray}
\Delta\rho|^{\rm High~order}_{\rm Conv} &\simeq& (^{+0.045}_{-0.018})\times 10^{-3},\\
\Delta\rho|^{\rm High~order}_{\rm PMC}  &\simeq& (\pm0.006)\times 10^{-3}.
\end{eqnarray}
It indicates that after applying the PMC, the precision of pQCD prediction can be greatly improved. By further taking the error caused by $\Delta\alpha_s(M_Z)=\pm0.0010$, which are $(\mp0.011)\times10^{-3}$ for both conventional and PMC scale-setting approaches, our final predictions of $\Delta\rho$ are
\begin{eqnarray}
\Delta\rho|_{\rm Conv} &\simeq& (8.211^{+0.046}_{-0.021})\times10^{-3},\\
\Delta\rho|_{\rm PMC}  &\simeq& (8.204\pm0.012)\times10^{-3}, \label{pmcresults}
\end{eqnarray}
where the error of PMC prediction is dominantly effected by $\Delta\alpha_s(M_Z)$. Thus a more precise measurement on the reference point $\alpha_s(M_Z)$ is helpful for a more precise pQCD prediction.

{\bf As a summary}, we have analyzed the QCD corrections to the electroweak $\rho$ parameter by using the PMC single-scale approach. The PMC eliminates the conventional renormalization scale ambiguity by applying the RGE to fix a scale-invariant effective coupling $\alpha_s(Q_*)$. It also greatly suppresses the residual scale dependence due to unknown higher-order terms, leads to a more convergent series, and then results in a more reliable and precise prediction, $\Delta\rho|_{\rm PMC}\simeq(8.204\pm0.012)\times10^{-3}$.

\begin{table*}[htb]
\centering
\begin{tabular}{c c c c c c c c c c c}
\hline
& \multicolumn{4}{c}{$\delta M_{W}|_{i}$ (MeV)} & \multicolumn{4}{c}{$\delta\sin^{2}\theta^{\rm left}_{\rm eff}|_{i}(\times10^{-5})$} \\
\cline{2-5}
\cline{5-9}
~~~ ~~~ & NLO & $\rm N^{2}LO$  & $\rm N^{3}LO$  & $\rm N^{4}LO$  & NLO  & $\rm N^{2}LO$  & $\rm N^{3}LO$  & $\rm N^{4}LO$ \\
\hline
Conv & $-51.62^{+4.07}_{-5.43}$ & $-10.12^{+3.49}_{-1.17}$  & $-2.64^{+1.20}_{-0.45}$ & $-0.69^{+0.38}_{-0.16}$ &$+28.64^{+3.02}_{-2.26}$ & $+5.62^{+0.65}_{-1.94}$  & $+1.47^{+0.25}_{-0.67}$ & $0.38^{+0.09}_{-0.21}$ \\

PMC & $-71.35$ & $+4.91$ & $+1.68$ & $-0.26$ & $39.59$ & $-2.73$ & $-0.93$ & $+0.14$ \\
\hline
\end{tabular}
\caption{The shifts $\delta M_{W}$ and $\delta\sin^{2}\theta^{\rm left}_{\rm eff}$ due to the QCD improved $\rho$ parameter before and after applying the PMC, where the conventional errors are caused by varying $\mu_r\in[M_t/2,2M_t]$ and the central value is for $\mu_r=M_t$. }
\label{table1}
\end{table*}

As an application, the $\Delta\rho$ parameter can be used to determine a variety of electroweak quantities such as the $W$-boson mass $M_W$ and the effective leptonic weak mixing angle $\rm sin^2\theta^{\rm lept}_{\rm eff}$, both of which could be measured with high precision. The recent world average of the uncertainties of $\rm sin^2\theta^{\rm lept}_{\rm eff}$ and $M_{W}$ are $\delta\sin^{2}\theta^{\rm lept}_{\rm eff} =33\times10^{-5}$ and $\delta M_{W} = 12$ MeV~\cite{PDG2020}. At the CEPC, the uncertainties could reach up to $\delta\sin^{2}\theta^{\rm lept}_{\rm eff}=1.0\times10^{-5}$ and $\delta M_{W} =1$ MeV~\cite{Liang:2019shd}. At the International Linear Collider (ILC), the uncertainties could reach up to $\delta\sin^{2}\theta^{\rm lept}_{\rm eff}=1.3\times10^{-5}$ and $\delta M_{W}=2.4$ MeV~\cite{deBlas:2019wgy, Fujii:2019zll}. At the Future Circular Colliders (FCC), the  uncertainties could reach up to $\delta\sin^{2}\theta^{\rm lept}_{\rm eff} =(0.2-0.5)\times10^{-5}$ and $\delta M_{W} =1$ MeV~\cite{Locci:2020knq}. Theoretically, the uncertainties of $M_W$ and $\rm sin^2\theta^{lept}_{eff}$ are related to the parameter $\Delta\rho$~\cite{Schroder:2005db}, and we present the derived $\delta M_{W}$ and $\delta\sin^{2}\theta^{\rm left}_{\rm eff}$ up to N$^4$LO-level in Table~\ref{table1}. The N$^4$LO predictions on the uncertainties are $\delta M_{W}|_{\rm N^4LO}=-0.26~{\rm MeV}$ and $\delta\sin^{2}\theta^{\rm left}_{\rm  eff}|_{\rm N^4LO}=0.14\times10^{-5}$, which are well below the present experimental precision. Thus we shall have great chances to test SM with high precision.

\hspace{1cm}

{\bf Acknowledgments:} We thank Sheng-Quan Wang for helpful discussions. This work was supported in part by the Chongqing Graduate Research and Innovation Foundation under Grant No.CYB21045 and No.ydstd1912, by the Natural Science Foundation of China under Grant No.11625520 and No.12047564, and by the Fundamental Research Funds for the Central Universities under Grant No.2020CQJQY-Z003. \\

\end{document}